\documentclass[proof]{WileyASNA-v1}
\usepackage{supertabular}
\usepackage{natbib}
\usepackage{lscape}
\articletype{Article Type}%

\received{29 January 2021}
\revised{20 July 2021}
\accepted{27 August 2021}

\begin{document}

\title{Identification of additional young nearby runaway stars based on Gaia data release 2 observations and the lithium test\protect\thanks{Based on observations obtained with telescopes of the University Observatory Jena, which is operated by the Astrophysical Institute of the Friedrich-Schiller University, and telescopes of the Fred L.\ Whipple Observatory, which is operated by the Harvard-Smithsonian Center for Astrophysics Cambridge MA.}}

\author[1]{Richard Bischoff*}
\author[1]{Markus Mugrauer}
\author[2]{Guillermo Torres}
\author[1]{Michael Geymeier}
\author[1]{Ralph Neuh\"auser}
\author[1]{Wolfgang Stenglein}
\author[1]{Kai-Uwe Michel}

\authormark{Bischoff \textsc{et al.}}

\address[1]{\orgdiv{Astrophysikalisches Institut und Universit\"{a}ts-Sternwarte Jena},  \orgaddress{\state{Schillerg\"{a}{\ss}chen 2, 07745 Jena}, \country{Germany}}}
\address[2]{\orgdiv{Center for Astrophysics $\vert$ Harvard \& Smithsonian}, \orgaddress{\state{60 Garden Street, Cambridge MA 02138}, \country{United States of America}}}

\corres{*R. Bischoff, Astrophysikalisches Institut und Universit\"{a}ts-Sternwarte Jena, Schillerg\"{a}{\ss}chen 2, 07745 Jena, Germany \email{richard.bischoff@uni-jena.de}}

\abstract{Runaway stars are characterised by their remarkably high space velocities, and the study of their formation mechanisms has attracted considerable interest. Young, nearby runaway stars are the most favorable for identifying their place of origin, and for searching for possible associated objects such as neutron stars. 
	
Usually the research field of runaway stars focuses on O- and B-type stars, because these objects are better detectable at larger distances than late-type stars. Early-type runaway stars have the advantage, that they evolve faster and can therefore better be confirmed to be young. In contrast to this, the catalogue of \textit{Young runaway stars within 3\,kpc} by \cite{tetzlaff} contains also stars of spectral type A and later. The objects in this catalogue were originally classified as young ($\le 50$ Myr) runaway stars by using \textit{Hipparcos} data to estimate the ages from their location in the \text{Hertzsprung-Russell} diagram and evolutionary models. 
	
In this article, we redetermine and/or constrain their ages not only by using the more precise second data release of the \textit{Gaia} mission, but also by measuring the equivalent width of the lithium (6708\,\AA) line, which is a youth indicator. Therefore, we searched for lithium absorption in the spectra of 51 target stars, taken at the University Observatory Jena between March and September 2020 with the \'Echelle spectrograph FLECHAS, and within additional TRES-spectra from the Fred L.\ Whipple Observatory. The main part of this campaign with its 308 reduced spectra, accessible at \texttt{VizieR}, was already published. In this work, which is the continuation and completion of the in 2015 initiated observing campaign, we found three additional young runaway star candidates.}

\keywords{stars: HRD, fundamental parameters; methods: observational, data analysis; techniques: spectroscopic; astronomical databases: catalogues}

\jnlcitation{\cname{%
\author{Bischoff, R.},
\author{Mugrauer, M.},
\author{Torres, G.}
et al.} (\cyear{2021}),
\ctitle{Identification of additional young nearby runaway stars based on Gaia data release 2 observations and the lithium test}, \cjournal{Astron.  Nachr.}, 2021}

\maketitle

\section{Introduction}\label{sec1}

Runaway stars can result from gravitational interactions in dense stellar clusters \citep{poveda} or from a supernova explosion in a binary system \citep{blaauw}. In both cases the ejected stars move on with higher velocities compared to typical field stars. Runaway stars can be traced back to their birth place by considering the influence of the Galactic potential. The correctness of those calculations depends mainly on the time since the ejection and will be even more challenging after several Myr. Credible runaway stars, that originated e.g. from the explosion in a binary system, should not be older than about 50\,Myr (including the lifetime of the progenitor star until the supernova).      
      
In 2011 the catalogue of \textit{Young runaway stars within 3\,kpc}\footnote{\url{http://cdsarc.u-strasbg.fr/viz-bin/cat/J/MNRAS/410/190}} was published by \citeauthor{tetzlaff}, which contains not only  O- and B-type stars, but also every type of star with a peculiar space velocity $v_{\text{pec}}>28$\,km/s. These objects were identified within a combined analysis of spatial, tangential and radial velocities, measured by the \textit{Hipparcos} satellite \citep{perryman}. Furthermore, their ages were derived by comparing luminosity and effective temperature to different evolutionary models.    

However, these age estimations can be improved and/or constrained with the more accurate second data release \citep{gaiadr2} of the \textit{Gaia} mission (\textit{Gaia}\,DR2) of the European Space Agency. Additionally, we searched for another youth indicator, namely the absorption line of the lithium doublet at 6708\,\AA\ based on \cite{neuhaeuser}, within our spectroscopic observing program for selected stars from the catalogue by \citeauthor{tetzlaff}. This observing program started in 2015 and the first results were already published in \cite{bischoff2020}. The remaining targets of this project are presented and discussed in this article. 

In section \ref{sec2}, we describe the sample selection, spectroscopic observations and data reduction. In section \ref{sec3}, we characterise the physical properties of our targets  based on their \textit{Gaia}\,DR2 data and in section \ref{sec4} we explain the measurements of the Li\,(6708\,\AA) line in all taken spectra. Section \ref{sec5} contains the age estimation of the dwarf stars. Finally, all results are discussed in section \ref{sec6} and we draw conclusions in section \ref{sec7}.

\section{Sample selection, Observations and data reduction}\label{sec2}

Our sample was selected from the catalogue by \cite{tetzlaff}. These targets had to be brighter than $V\leq8.5$\,mag in order to record spectra with sufficiently high signal-to-noise-ratio (SNR) $>50$ within integration times of a few minutes. Furthermore, their declination angle had to be $Dec>-14^{\circ}$ so that they can be observed at air masses $X<2.4$ from Jena. As mentioned in \cite{bischoff2020}, in total 460 stars were identified that fulfill these conditions and 308 of them were already observed and published. However, the remaining list of 152 stars could be further shortened with a rough analysis by using data from \textit{Gaia}\,DR2 and the catalogue of \cite{bailerjones} to rule out most of the giants. This results in a target list that contains 51 stars, which were observed with the fibre-linked \'{E}chelle spectrograph FLECHAS \citep{mugrauer2014} and processed in the following.         

Our spectra were taken with FLECHAS, operated at the Nasmyth-focus of the 90\,cm-telescope ($f/D=15$) of the University Observatory Jena \citep{pfau}. The observations, that include 153 spectra with a total integration time of $25.75$\,h, were carried out between March and September 2020.    

All spectra were recorded with the 1x1 binning mode of the spectrograph FLECHAS, using individual detector integration times in the range between 150\,s and 1200\,s dependent on the target brightness. The instrument has a resolving power of $R\approx9,300$ and covers a spectral range from  3900\,\AA\,\,to 8100\,\AA\,\,within 29 orders \citep{mugrauer2014}. Three spectra per target were always taken to remove cosmics and to reach a sufficiently high SNR, which was measured in all fully reduced spectra at $\lambda=6700$\,\AA, which is the centre of the spectral order with the Li\,(6708\,\AA) line. On average $\text{SNR}=101$ is reached in the FLECHAS spectra of our targets, with range from $50$ for HIP\,30030 to 201 for HIP\,19587. Further details are given in the observation log in Table\,\ref{tab:obslog}\hspace{-2mm}.    

Three flat-field frames of a tungsten lamp and three spectra of a thorium-argon (ThAr) lamp are recorded immediately before the observation of each target for calibration purposes. Each calibration file has an individual integration time of 5\,s. About 700 detected emission lines are available in the ThAr spectra for wavelength calibration. The long-term stability of the wavelength calibration of FLECHAS was confirmed in studies by \cite{irrgang}, \cite{bischoff}, \cite{heyne} and \cite{bischoff2020b}. Additionally, for the dark subtraction, three dark frames for all used integration times were taken in every observing night. An overscan region is always read out to measure and later correct the bias level. The FLECHAS detector has a typical read-noise of about 11\,$e^{-}$ and the gain is $1.3\,e^{-}$/ADU. The FLECHAS CCD-sensor and the whole instrument is described in detail in \cite{mugrauer2014}.

The observations were reduced with a dedicated pipeline for FLECHAS, developed at the Astrophysical Institute Jena, which includes dark and bias subtraction, flat-fielding, extraction and wavelength calibration of the individual spectral orders. Furthermore, including the final averaging and normalisation of the spectra \citep{mugrauer2014}. 

As part of a separate long-term spectroscopic monitoring program to measure radial velocities and discover binary systems in another sample of runaway stars from \cite{tetzlaff}, HIP\,2710 and HIP\,12297 were also observed between September 2013 and March 2017 with the Tillinghast Reflector Echelle Spectrograph TRES \citep{Szentgyorgyi2007, furesz2008}. The spectrograph is attached to the 1.5\,m Tillinghast reflector at the Fred L.\ Whipple Observatory on Mount Hopkins (Arizona, USA). This bench-mounted, fiber-fed instrument generates spectra at a resolving power of $R \approx 44,000$ that cover the wavelength region between 3800\,\AA\ and 9100\,\AA\ in 51 orders. Exposure times ranged from 60\,s to 250\,s, depending on brightness and weather conditions. Exposures of a ThAr lamp were taken before and after each science frame, and the observations were also reduced with a dedicated pipeline, which follows the procedure described in \cite{furesz2008}.    

\section{Target characterisation with Gaia\,DR2 data}\label{sec3}

The detailed characterisation of our targets in this article focusses mainly on data from the \textit{Gaia}\,DR2. We considered only gold flag photometry (as described by \citealt{andrae2018}) entries from the \textit{Gaia}\,DR2. The apparent brightness in the $G$-band for each target was corrected according to the brightness relations in \cite{weiler} and taking into account the new defined transmission profiles for the \textit{Gaia} filters from \cite{weiler2}. We did not use the estimates for the $G$-band extinction in \textit{Gaia}\,DR2, because they were not available for 16 targets of our sample and sometimes they were significantly overestimated, e.g. in the case of HIP\,56770, an extinction of $A_{\text{G}}=0.995_{-0.314}^{+0.188}$\,mag seems unrealistic, given its distance of $47.8\pm0.5$\,pc from \cite{bailerjones}. Therefore, we take extinctions in the $V$-band from the catalogue of \cite{gontcharov} instead and convert these values with relations from \cite{wang} to $G$-band extinctions.

Based on apparent brightness $G$ and extinction $A_{\text{G}}$ in the $G$-band, together with the distances $d$ from \cite{bailerjones}, we calculated the absolute $G$-band brightness $M_{\text{G}}$. The used values and results are presented in Table \ref{tab:gaia_infos}\hspace{-2mm}. Additionally, we list effective temperatures $T_{\text{eff}}$, stellar radii $R$ and luminosities $L$ of our targets, if available and the target was not a spectroscopic binary.

\begin{figure}[h!]
\centering\includegraphics[width=8.25cm,height=8.25cm,keepaspectratio]{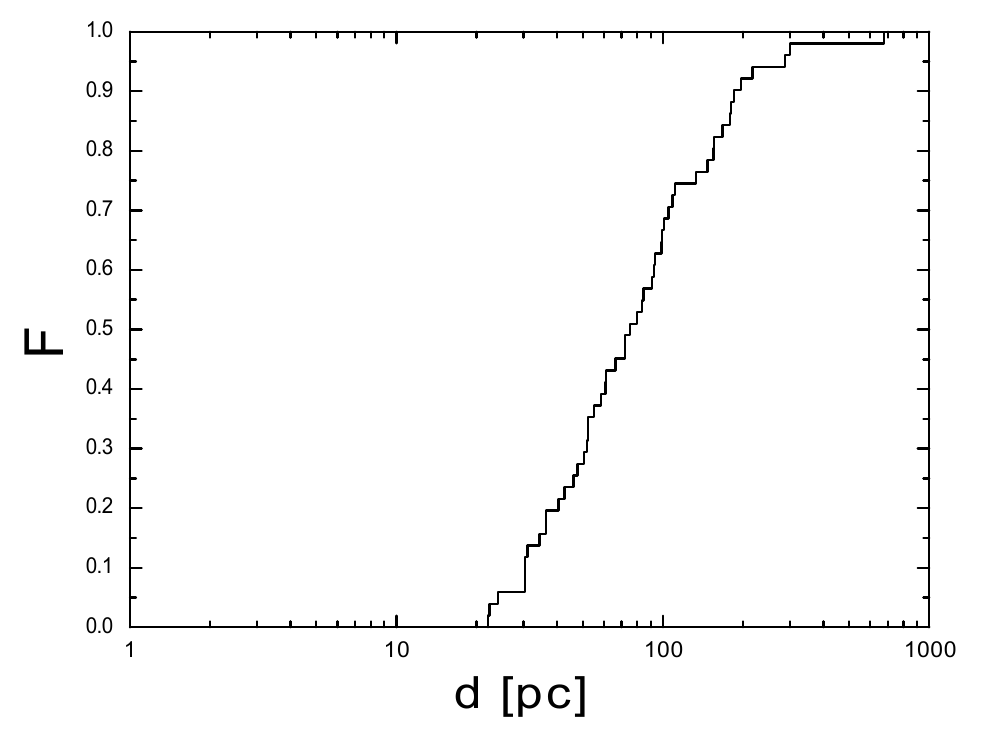}
\caption{The cumulative distribution function of the distances of our sample.}
\label{fig:dist}
\end{figure}

We show the distance distribution of our targets in Figure\,\ref{fig:dist}\hspace{-2mm}. It has a median distance of about 75.4\,pc. The individual distances range between $22.1_{-0.1}^{+0.1}$\,pc (HIP\,19855) and $676_{-290}^{+2027}$\,pc (HIP\,113811).     

However, not all needed parameters were available in \textit{Gaia}\,DR2, \cite{gontcharov} and \cite{bailerjones} for our complete target sample.
HIP\,57529, HIP\,106053, HIP\,113811 and HIP\,115906 were missing distance information and therefore, we had to calculate it from the \textit{Hipparcos} parallax \citep{vanleeuwen} instead. For the same four targets, we converted the apparent magnitude of the \textit{Hipparcos} system $H_{\text{p}}$ and the $V-I$ magnitudes, provided by the \textit{Hipparcos} catalogue \citep{perryman}, into the $G$-band with the relations by \cite{evans}. Their $T_{\text{eff}}$ were derived from their \textit{Hipparcos} spectral type (SpT) with the corresponding SpT-$\log(T_{\text{eff}})$-relations by \cite{damiani}. HIP\,113811 had also no entry in the extinction catalogue of \cite{gontcharov}. Therefore, we used the reddening $E(g-r)$ from \cite{green}, which was transformed into $A_{\text{r}}$ with the relation given there and afterwards into $A_{\text{G}}$ \citep{wang}.

We searched for multiplicity within our sample in the \textit{9th Catalogue of Spectroscopic Binary Orbits} \citep{pourbaix} to correct their absolute brightness. In the case of a double-lined spectroscopic binary, we took the mass ratio of the secondary and the primary component and converted it into a luminosity ratio via $L\propto M^{4.5}$, which is applicable for stellar masses between $2\,\text{M}_{\odot}$ and $0.5\,\text{M}_{\odot}$ \citep{salaris}, suitable for our sample of spectral types ranging between A7 and K2. The luminosity ratios were then used to determine how much brighter is the binary in comparison to a single source. We assumed that the brightness difference between the secondary and the primary is at least 1\,mag for the single-lined binaries, if no further information about mass ratios or the systems were available. It follows that those systems could be up to $\sim0.364$\,mag brighter than a corresponding single star. The identified spectroscopic binaries and their mass ratios are listed in Table\,\ref{specbinmass}\hspace{-2mm}.  

\begin{table}[h!]
\caption{The spectroscopic binary stars within our target sample. We list their mass ratio $M_{2}/M_{1}$, if available, and the corresponding reference.}
\centering
\begin{tabular}{lcl}
\hline
 Target & $M_{2}/M_{1}$  & ref. \\
			\hline
HIP\,5081	& $0.902\pm0.009$   & a  \\
HIP\,22524	& -   &  \\
HIP\,26690	& $0.938\pm0.049$   & b \\
HIP\,64312	& $0.729\pm0.006$   & c \\
HIP\,71631	& $0.556\pm0.127$   & d \\
HIP\,82798	& -   &  \\
HIP\,112821	& -   &  \\
HIP\,114379	& $0.973\pm0.001$   & e \\
\hline
\end{tabular} 
\newline
\begin{flushleft}
a \cite{griffin}\\
b \cite{nordstrom}  \\
c \cite{escorza} \\
d \cite{koenig} \\
e \cite{fekel}\\
\end{flushleft}                              		
\label{specbinmass}
\end{table}

\begin{figure*}[h!]
\centering\includegraphics[width=17.5cm,height=17.5cm,keepaspectratio]{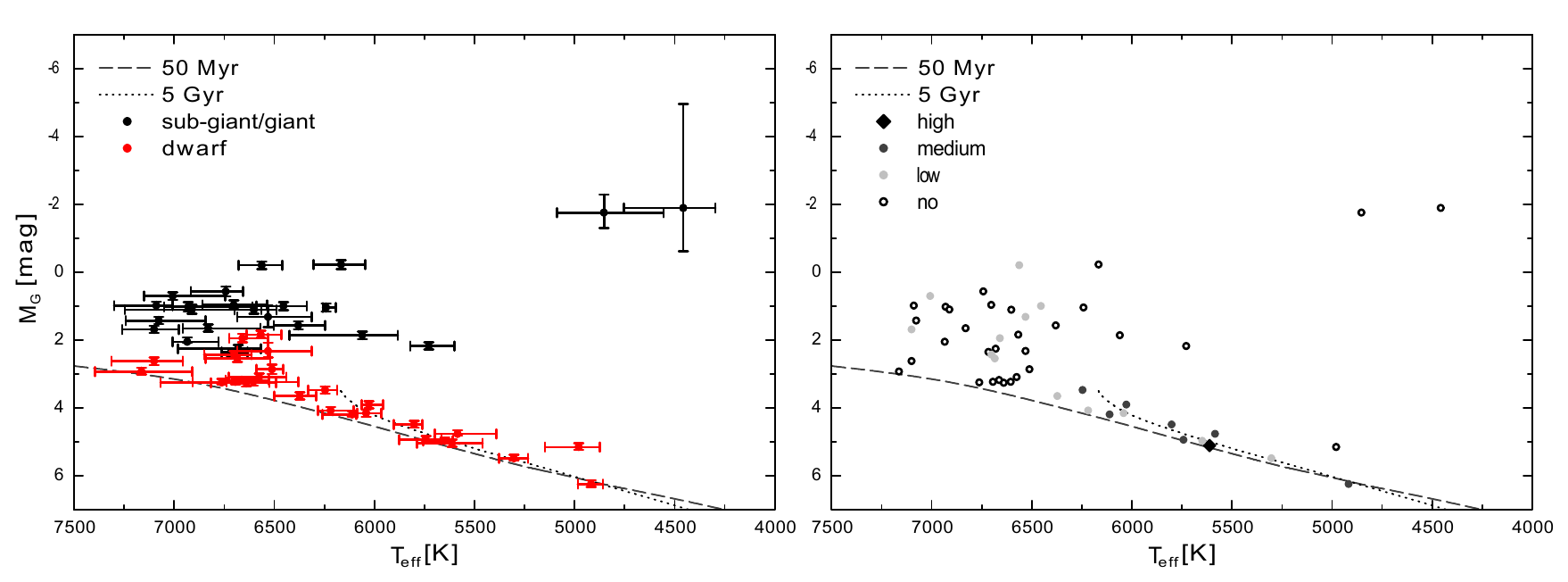}
\caption{\text{Hertzsprung-Russell} diagram with all 51 observed targets. On the \textbf{left}, we show sub-giant/giant stars marked with full black dots and dwarf stars with full red dots. On the \textbf{right}, the targets are characterised by the measured equivalent width of the Li\,(6708\,\AA) line: Targets with $EW_{\text{Li}}< 3\cdot \sigma EW_{\text{Li}}$ have \textit{no} significant lithium detection, while objects with $EW_{\text{Li}}\geq 3\cdot\sigma EW_{\text{Li}}$ and $EW_{\text{Li}}<100$\,m\AA\,\,are ranked as \textit{low}, 100\,m\AA\,\,$\leq\,EW_{\text{Li}}<200$\,m\AA\,\,as \textit{medium} and $EW_{\text{Li}}\geq200$\,m\AA\,\,as \textit{high}. We have also plotted the PARSEC isochrones \citep{bressan} for 50\,Myr and 5\,Gyr with solar metallicity $Z=0.0152$ in both distributions.}
\label{fig:HRD}
\end{figure*}

Our targets are illustrated in a \text{Hertzsprung-Russell-Diagram} (HRD) in Figure\,\ref{fig:HRD}\hspace{-2mm}. For example, HIP\,113811 ($M_{\text{G}}=-1.891_{-3.070}^{+1.275}$\,mag) and HIP\,115906 ($M_{\text{G}}=-1.754_{-0.531}^{+0.459}$\,mag) can be excluded as possible young runaway stars, because they are far too bright for their given $T_{\text{eff}}$ to be dwarf stars and are clearly located on the giant branch. Typical $M_{\text{G}}$-$T_{\text{eff}}$-relations for dwarf stars are given in \cite{mamajek}\footnote{\url{https://www.pas.rochester.edu/~emamajek/EEM_dwarf_UBVIJHK_colors_Teff.txt}} and \cite{baraffe}\footnote{\url{http://perso.ens-lyon.fr/isabelle.baraffe/BHAC15dir/}}. However, it is not always easy to decide, based on their location in the HRD alone, whether an object is either a pre-main-sequence star or it has already left the main-sequence. Therefore, we studied their listed surface gravities in the \textit{StarHorse} catalogue \citep{anders} and if the given range of $\log(g[\text{cm/s}^{2}]) \gtrsim 3.8$, the target was classified as dwarf star.             

To identify the young stars among the dwarfs, further analysis is needed.

\section{Li\,(6708\,\AA) Equivalent width measurements and abundances}\label{sec4}

The Ca\,(6718\,\AA) line was used to correct the doppler shift in every spectrum, adopting $\lambda_0 = 6717.685$\,\AA\,\,as laboratory wavelength as listed in the ILLSS catalogue \citep{coluzzi}, because it is the most prominent spectral line nearby Li\,(6708\,\AA), and is also detected in the same spectral order. 

We measured the equivalent width via a direct integration of the line profiles in the reduced spectra by using the IRAF \citep{tody} task \texttt{splot} as explained in \cite{bischoff2020}. 

The equivalent widths of the Li\,(6708\,\AA) line of all targets are given in Table\,\ref{tab:Li_all}\hspace{-2mm}. Furthermore, in appendix \textit{\nameref{app2}} and \textit{\nameref{app3}}, we illustrate all FLECHAS spectra that show a significant detection, that means $EW_{\text{Li}}\geq 3\cdot \sigma EW_{\text{Li}}$. All spectra are sorted by their spectral type according to SpT-$\log(T_{\text{eff}})$-relations from \cite{damiani}. The uncertainty of the spectral classification is about two sub-classes. In additional TRES-spectra of HIP\,2710 and HIP\,12297, we searched for lithium as described above. The measured average equivalent width from the four spectra of HIP\,2710 is $EW_{\text{Li}}=(43\pm7)$\,m\AA\,. In the case of HIP\,12297, none of the 24 spectra showed a significant Li\,(6708\,\AA) line. These results are consistent with the measurements from FLECHAS.\\

Our equivalent widths measurements of the identified dwarf stars with significant lithium detection were converted into abundances by using curves of growth from \cite{soderblom}, that are based on an abundance scale of $\log_{10}(N)_{\text{H}}=12$ and we assigned the best matching values of $\log_{10}(EW)$ and $T_{\text{eff}}$ (Table\,2 in \citealt{soderblom}) to those of our sample. Indeed, some of our $T_{\text{eff}}$ values were outside the covered range of \cite{soderblom}. Therefore, we fit quadratic polynomials as a function of $T_{\text{eff}}$ for constant values of $\log_{10}(EW)$. The results of this conversion are presented in Table\,\ref{abundance}\hspace{-2mm}.

\begin{table}[h!]
\caption{The classified dwarf stars with their identification numbers as shown in Figure\,\ref{fig:HRD}\hspace{-2mm}, with their effective temperatures $T_{\text{eff}}$ from \textit{Gaia}\,DR2 and the measured equivalent widths of the Li\,(6708\,\AA) line $EW_{\text{Li}}$. We list only targets with significant lithium detection. The abundances $\log_{10}(N_{\text{Li}})$ based on \cite{soderblom} are given in the last column.}
\centering
\resizebox{0.5\textwidth}{!}{
\begin{tabular}{clccc}
\hline
		id. nr. & Target & $T_{\text{eff}}$ [K] & $EW_{\text{Li}}$ [m\AA] & $\log(N_{\text{Li}})$   \\
\hline
1  & HIP\,28469		& $6701_{-87}^{+150}$    & $~~61\pm12$  & $3.196_{-0.318}^{+0.072}$ \\
2  & HIP\,113174	& $6684_{-164}^{+159}$   & $~~57\pm11$  & $3.135_{-0.318}^{+0.133}$ \\
3  & HIP\,26690 	& $6659_{-128}^{+67}$    & $~~66\pm15$  & $3.196_{-0.318}^{+0.143}$ \\
4  & HIP\,2710 		& $6372_{-81}^{+129}$    & $~~36\pm12$  & $2.519_{-0.152}^{+0.359}$ \\
5  & HIP\,22524     & $6246_{-60}^{+86}$     & $119\pm12$   & $3.260_{-0.090}^{+0.090}$ \\
6  & HIP\,54531 	& $6218_{-114}^{+63}$    & $~~85\pm14$  & $3.011_{-0.138}^{+0.079}$ \\
7  & HIP\,51386 	& $6111_{-21}^{+147}$    & $120\pm11$   & $3.043_{-0.087}^{+0.217}$ \\
8  & HIP\,44212 	& $6041_{-74}^{+44}$     & $~~64\pm13$  & $2.593_{-0.120}^{+0.136}$ \\
9  & HIP\,30030 	& $6027_{-69}^{+36}$     & $190\pm24$   & $3.517_{-0.271}^{+0.177}$ \\
10  & HIP\,114385 	& $5801_{-41}^{+103}$    & $111\pm10$   & $2.726_{-0.082}^{+0.317}$ \\
11  & HIP\,115527   & $5742_{-78}^{+135}$    & $126\pm13$   & $2.808_{-0.082}^{+0.332}$ \\
12  & HIP\,19855    & $5648_{-39}^{+108}$    & $~~71\pm12$  & $2.432_{-0.372}^{+0.069}$  \\
13  & HIP\,16563 	& $5612_{-153}^{+176}$   & $254\pm14$   & $3.367_{-0.522}^{+0.522}$ \\
14  & HIP\,71631    & $5584_{-193}^{+115}$   & $197\pm13$   & $3.000_{-0.132}^{+0.271}$ \\
15  & HIP\,7576     & $5303_{-71}^{+73}$     & $114\pm12$   & $2.196_{-0.073}^{+0.361}$  \\
16  & HIP\,40774 	& $4917_{-58}^{+67}$     & $119\pm12$   & $1.980_{-0.400}^{+0.400}$ \\
\hline
\end{tabular}
}                                  		
\label{abundance}
\end{table}

\section{Age estimation}\label{sec5}

\begin{figure}[h!]
\centering\includegraphics[width=8.3cm,height=8.25cm,keepaspectratio]{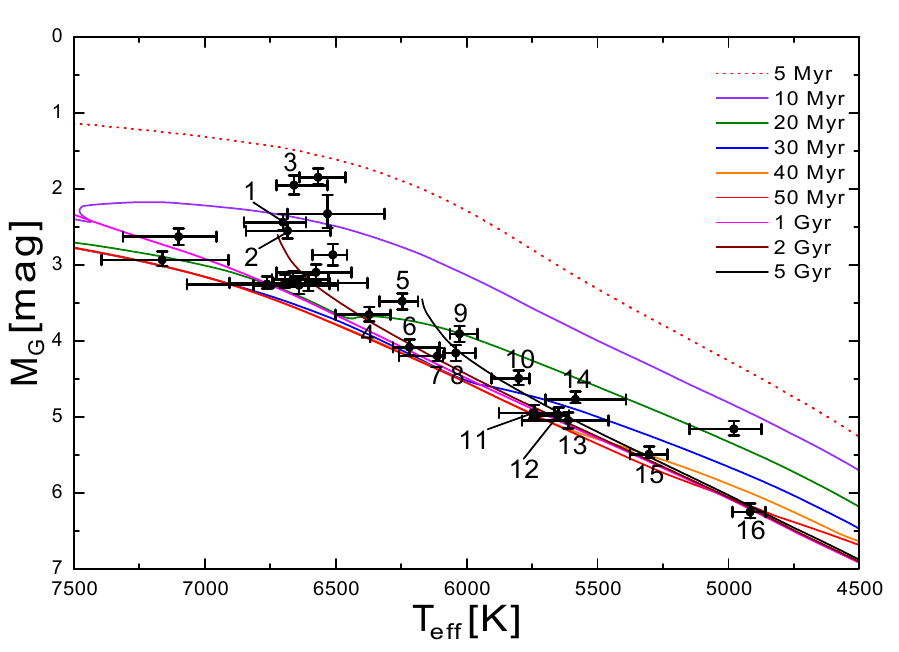}
\caption{The dwarf stars of our sample are illustrated in the \text{Hertzsprung-Russell} diagram with different isochrones, using the stellar evolutionary models of \cite{bressan} for solar metallicity $Z=0.0152$.}
\label{fig:zwerge}
\end{figure}

\begin{figure}[h!]
\centering\includegraphics[width=8.3cm,height=8.25cm,keepaspectratio]{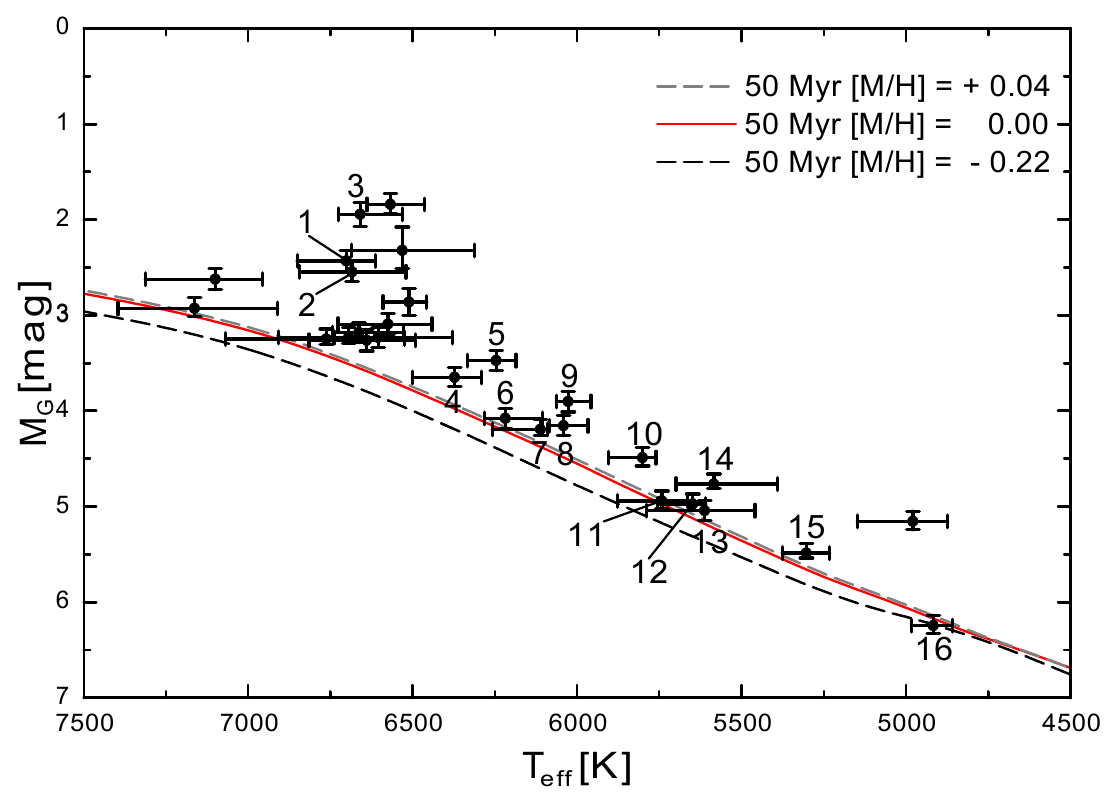}
\caption{The dwarf stars of our sample plotted in the \text{Hertzsprung-Russell} diagram with 50\,Myr isochrones for different metallicities $[\text{M}/\text{H}]$ using the stellar evolutionary models of \cite{bressan}.}
\label{fig:metal}
\end{figure}

We can constrain the ages of our identified dwarf stars with further isochrones. The isochrones in Figure\,\ref{fig:zwerge} were calculated with models of \cite{bressan} for metallicity $Z=0.0152$. Assuming solar metallicity is justified, because all dwarfs exhibit an average metallicity of $[\text{M}/\text{H}]=0.09$ with a standard deviation of $0.13$\,dex. This estimate is based on a compilation of metallicities for our stars from the \texttt{VizieR} database \citep{ochsenbein}, taken from the catalogues by \cite{brewer}, \cite{casagrande}, \cite{casamiquela}, \cite{franchini}, \cite{gray2003}, \cite{gray2006}, \cite{kunder}, \cite{luck}, \cite{mann}, \cite{marsden}, \cite{petigura}, \cite{strassun} and \cite{valenti}. The influence of the metallicity scatter is shown in Figure\,\ref{fig:metal}\hspace{-2mm}. Here, we show as an example the 50\,Myr isochrone. The differences between different metallicities are smaller or at least comparable with the uncertainties of effective temperature or absolute brightness. Hence, they do not effect the age estimation significantly.             

However, we have to consider that many of our dwarf stars are consistent with more than one isochrone within their uncertainties in Figure\,\ref{fig:zwerge}\hspace{-2mm}, especially if they are matching one of the Gyr isochrones. For that reason, the estimated ages based on the location in the HRD in Table\,\ref{clearname}\hspace{-2mm}, are sometimes only listed with lower limits. Additional information, as explained in the following, are necessary to identify and/or further constrain the young stars among our targets. 

\begin{figure}[h!]
\centering\includegraphics[width=8.75cm,height=8.75cm,keepaspectratio]{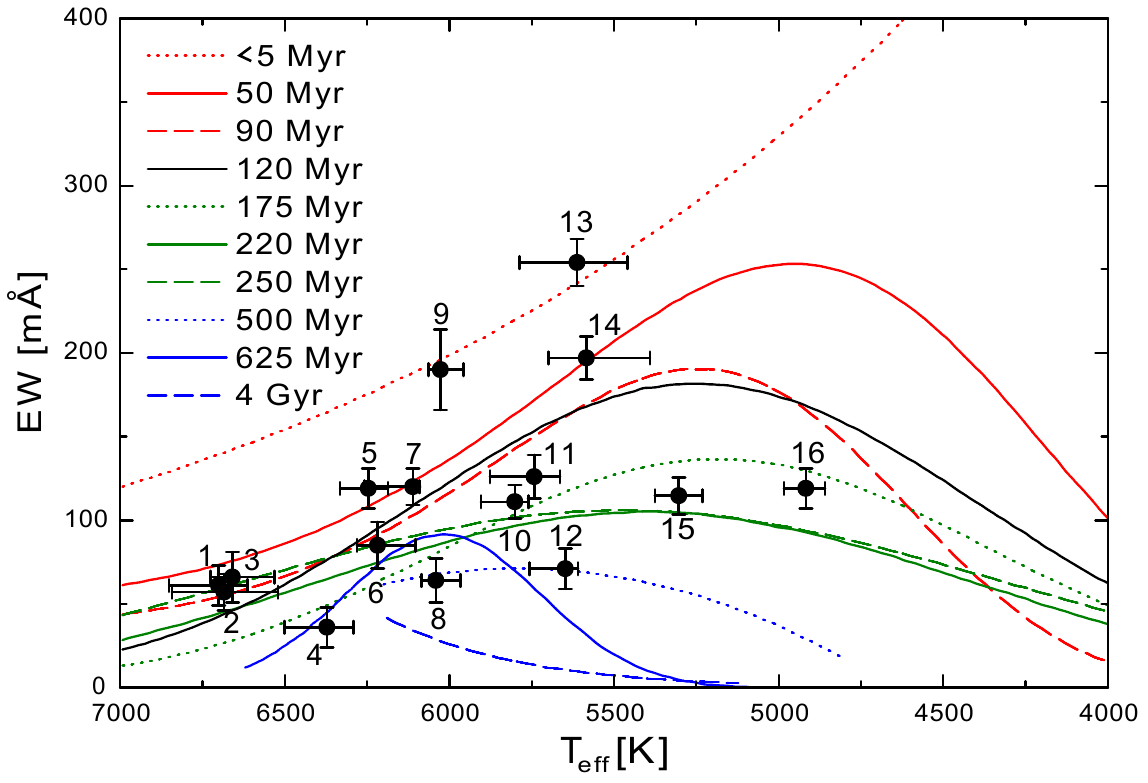}
\caption{Lithium as age indicator for our 16 youngest F\,G\,K\,M stars. We show the equivalent widths of the Li\,(6708\,\AA) line dependant on the effective temperature for different ages, that are based on fits done by Eric Mamajek. The derived ages are given in Table\,\ref{clearname}\hspace{-2mm}.}
\label{fig:Li}
\end{figure}

We compared all dwarf stars with significant lithium detection in their spectra to distributions of stellar clusters with known ages, as illustrated in Figure\,\ref{fig:Li}\hspace{-2mm}. The curves are polynomial fits for observed average equivalent width measurements dependent on $T_{\text{eff}}$ (cluster data and fits from E. Mamajek, priv. communication). E. Mamajek`s original plot is available online\footnote{\url{http://www.pas.rochester.edu/~emamajek/images/li.jpg}}.       

The age errors for this method appear to be $10\%-20\%$, as stated by \cite{soderblom2014}, and the detection of lithium in a low-mass star with known effective temperature can give an upper limit to its age. HIP\,22524 (\#\,5 in Table\,\ref{clearname} and Figure\,\ref{fig:Li}\hspace{-2mm}) is only consistent with the 50\,Myr curve and within its uncertainties it reaches clearly the area for stars that are younger than 50\,Myr. Hence, its assigned age is $\le50$\,Myr. HIP\,51386 (\#\,7) and HIP\,71631 (\#\,14) fit with more than one age curve and are also consistent with ages below 50\,Myr within their uncertainties. Therefore, they were considered to be $\le50\,...\,120$\,Myr and $\le50\,...\,90$\,Myr, respectively.   

HIP\,30030 (\#\,9) and HIP\,16563 (\#\,13) should be handled with care, because as stated by \cite{soderblom2014} the lithium method does not give reliable age estimations below 20\,Myr. For that reason, these stars were classified to have an age of $<50$\,Myr. Furthermore, for $T_{\text{eff}}>6300$\,K the $<5$\,Myr age curve in Figure\,\ref{fig:Li}\hspace{-2mm} is an extrapolation, because in this range no stars with lithium and the corresponding age were observed.

The remaining dwarfs with a significant detected Li\,(6708\,\AA) line cross more than one age curve in Figure\,\ref{fig:Li}\hspace{-2mm}. Their age estimation was encircled by the youngest age curve and the oldest age curve, that were hit within their uncertainties. For example, HIP\,19855 (\#\,12) matches the curves for 500\,Myr and 625\,Myr and therefore, it was estimated to have an age of 500 to 625\,Myr. The age estimations for the others stars are given in the corresponding column of Table\,\ref{clearname}\hspace{-2mm}.

HIP\,40774 (\#\,16) is about 175\,Myr old, because it only fit with those age curve.

\begin{table*}[h!]
\caption{Our dwarf stars with their identification numbers as shown in Figure\,\ref{fig:zwerge}\hspace{-2mm}, Figure\,\ref{fig:metal} and Figure\,\ref{fig:Li}\hspace{-2mm}, listed with their SpT derived from their $T_{\text{eff}}$ using the SpT-$\log(T_{\text{eff}})$-relation from \cite{damiani}, their distances $d$ according to \cite{bailerjones}, the measured equivalent width of the Hydrogen Balmer line $EW_{\text{H}\alpha}$ (all in absorption) and the measured equivalent width of Li\,(6708\,\AA) $EW_{\text{Li}}$.
Furthermore, we list the estimated age derived from the position in the HRD, as well as the age according to the lithium test from this work and the age from \cite{tetzlaff}.}
\centering
\resizebox{1.0\textwidth}{!}{
\begin{tabular}{clccccccc}
\hline
id. nr. & Target & SpT  & $d$ [pc] &   $EW_{\text{H}\alpha}$ [m\AA]  & $EW_{\text{Li}}$ [m\AA] & age (HRD) [Myr]   & age ($EW_{\text{Li}}$) [Myr] & age [Myr]  \\
	 		 & &  $\ast$    & Bailer-Jones  & $\ast$      &   $\ast$   & $\ast$  & $\ast$   & Tetzlaff \\
\hline
  & HIP\,116983	& A7\,V     & $111\pm1$ 	  & $1336\pm21$	& ~~- 		  & $>10, <2000$   		& -       			& $~~41.9\pm19.1$  \\
  & HIP\,111647 & A7\,V     & $~~84.6\pm0.4$  & $1173\pm21$ & ~~- 		  & $>10, <2000$		& -     			& $~~41.9\pm32.4$  \\
  & HIP\,47725 	& F2\,V     & $~~72.1\pm0.2$  & $1185\pm19$ & ~~-		  & $\ge20, <5000 $ 	& -         		& $~~41.9\pm31.9$  \\
1 & HIP\,28469 	& F2\,V     & $~~91.1\pm0.4$  & $1332\pm18$ & ~~$61\pm12$ & $\ge10 $			&$\le50\,...\,220$ 	& ~~- \\
  & HIP\,117835 & F2\,V     & $~~83.7\pm0.3$  & $1397\pm19$ & ~~-         & $>10, <5000$		& -               	& $~~41.9\pm32.4$    \\
2 & HIP\,113174 & F2\,V     & $~~42.8\pm0.1$  & $1317\pm17$ & ~~$57\pm11$ & $\ge10, \leq2000$   &$\le50\,...\,220$ 	& ~~-\\
  & HIP\,25749 	& F2\,V     & $~~55.1\pm0.1$  & $1249\pm19$ & ~~- 		  & $>10, <5000$		& -          		& $~~41.9\pm32.4$  \\
3 & HIP\,26690 	& F2\,V     & $167\pm2$  	  & $1324\pm22$ & ~~$66\pm15$ & $>5,<10$  			&$\le50\,...\,220$ 	& $~~35.8\pm21.9$\\
  & HIP\,111446 & F5\,V     & $~~66.4\pm0.4$  & $1321\pm19$ & ~~- 		  & $>10, <5000$   		& -         		& $35.8\pm8.8$\\
  & HIP\,5913 	& F5\,V     & $~~58.6\pm0.2$  & $1134\pm18$ & ~~-    	  & $>10, <5000$   		& -  				& $~~41.9\pm32.4$   \\
  & HIP\,49700  & F5\,V     & $~~80.0\pm0.4$  & $1242\pm14$ & ~~-         & $>10, <5000$   	& -              	& $~~41.9\pm32.4$   \\
  & HIP\,5081   & F5\,V & $~~52.5\pm0.4$  & $1038\pm20; 567\pm23^{\dagger}$ & ~~-& $>5, <10$ 	& -         		& ~~-  \\
  & HIP\,57529 	& F5\,V     & $~~92.5\pm6.3$  & $1133\pm12$ & ~~- 		  & $>5,<20$			& - 				& ~~-\\
  & HIP\,35219	& F5\,V     & $~~52.2\pm1.0$  & $1134\pm17$ & ~~- 		  & $>10, <5000$		& -     			& ~~-  \\
4 & HIP\,2710 	& F6\,V     & $~~40.6\pm0.1$  & $1344\pm18$ & ~~$36\pm12$ & $\ge20, <5000$		&$175\,...\,625$  	& $17.8\pm5.6$  \\
5 & HIP\,22524 	& F6\,V     & $~~50.4\pm0.1$  & $1225\pm15$ & $119\pm12$  & $>10, <5000$		&$\le50$         	& $17.1\pm3.0$  \\
6 & HIP\,54531 	& F6\,V     & $~~61.0\pm0.2$  & $1065\pm21$ & ~~$85\pm14$ & $>20, <5000$   		&$90\,...\,625$ 	& $~~38.8\pm22.9$  \\
7 & HIP\,51386  & F8\,V     & $~~31.0\pm0.1$  & $~~959\pm14$& $120\pm11$  & $>20, <5000$   		&$\le50\,...\,120$ 	& $25.5\pm4.3$    \\
8 & HIP\,44212 	& F8\,V     & $~~46.2\pm0.1$  & $1182\pm19$ & ~~$64\pm13$ & $>20$   			&$175\,...\,500$  	& $39.8\pm4.6$\\
9 & HIP\,30030 	& F8\,V     & $~~51.9\pm0.1$  & $~~722\pm30$& $190\pm24$  & $\ge20$   			&$<50$          	& $26.6\pm5.3$  \\
10& HIP\,114385 & G0\,V     & $~~30.3\pm0.1$  & $~~918\pm14$& $111\pm10$  & $>20$   			&$175\,...\,250$  	& $~~35.1\pm10.5$\\
11& HIP\,115527 & G0\,V     & $~~30.4\pm0.1$  & $~~838\pm16$& $126\pm13$  & $>30$   			&$90\,...\,175$ 	& $35.8\pm9.7$\\
12& HIP\,19855 	& G2\,V     & $~~22.1\pm0.1$  & $~~968\pm15$& ~~$71\pm12$ & $>30$  				&$500\,...\,625$  	& $~~39.5\pm17.4$   \\
13& HIP\,16563  & G2\,V     & $~~36.4\pm0.1$  & $~~543\pm18$& $254\pm14$  & $>20$   			&$<50$          	& $16.7\pm1.1$   \\
14& HIP\,71631  & G5\,V     & $~~34.4\pm0.1$  & $~~813\pm20$& $197\pm13$  & $\ge20, \leq30$  	&$\le50\,...\,90$   & $27.6\pm4.2$    \\
15& HIP\,7576 	& G8\,V     & $~~24.0\pm0.1$  & $~~933\pm12$& $114\pm12$  & $>30$   			&$220\,...\,250$ 	& $40.7\pm7.9$\\
  & HIP\,114379 & K2\,V     & $~~30.4\pm0.1$  & $274\pm17; 43\pm8^{\dagger}$  & ~~-         & $>10, <30$   &-    & $~~23.9\pm11.2$   \\
16& HIP\,40774 	& K2\,V     & $~~22.4\pm0.1$  & $~~703\pm14$& $119\pm12$  & $>40$   			&$\sim175$ 			& $~~54.9\pm10.4$\\
\hline
\end{tabular}
    }
\begin{flushleft}
$^{\ast}$ this work\\
$^{\dagger}$ both H$\alpha$-lines of this spectroscopic binary could be measured
\end{flushleft}                              		
\label{clearname}
\end{table*}

\section{Discussion}\label{sec6}

The aim of our project was to identify and/or confirm young mid- and late-type runaway stars from the catalogue by \cite{tetzlaff} based on their location in the HRD with the more accurate \textit{Gaia}\,DR2 data. In addition, we took spectra and searched for the absorption of the Li\,(6708\,\AA) line, which is a youth indicator. Our sample consists of 2 A-type, 38 F-type, 7 G-type and 4 K-type stars. Their SpT was assigned based on their $T_{\text{eff}}$ with the SpT-$\log(T_{\text{eff}})$-relation from \cite{damiani}.     

We studied the surface gravity of our targets within the \textit{StarHorse} catalogue to rule out possible sub-giants, that have typically $\log(g[\text{cm/s}^{2}]) < 3.8$. As a result of this, 23 targets could be excluded as already evolved stars.   
 
The main goal of this study was to find stars that are younger or about 50\,Myr. Therefore, we used isochrones, as illustrated in Fig.\,\ref{fig:zwerge}\hspace{-2mm}, to give an age limit of our identified dwarf stars. Due to their derived scatter of metallicity as explained above, assuming solar metallicity for our sample and using it for isochrone fitting seems reasonable. We considered possible multiplicity within our sample and checked the catalogue of \cite{pourbaix} for spectroscopic binaries, to correct their position in the HRD. We list all identified binaries in Table\,\ref{RV}\hspace{-2mm} with their measured radial velocity, which was determined from the Ca\,(6718\,\AA) line. The secondary component of the double-lined binaries HIP\,5081 and HIP\,114379 could also be measured.\\  

Nearly all dwarf stars are consistent within their uncertainties with isochrones in the range of a few Gyr. Therefore, another indicator is needed to confirm the youth of the targets. For this, we measured the equivalent width of the Li\,(6708\,\AA) line. HIP\,16563 has the strongest lithium line with $(254\pm14)$\,m\AA\,. In contrast to this, 30 stars of the sample showed no significant Li\,(6708\,\AA) line within their spectra.  

Equivalent width measurements of the dwarfs were then converted into abundances using the curves of growth from \cite{soderblom}. HIP\,44212 is also listed in the catalogues of \citep{ramirez} and \citep{lambert} and their measurements ($\log(N_{\text{Li}})=2.65\pm0.03$ and $\log(N_{\text{Li}})=2.55\pm0.10$, respectively) are consistent with our determined lithium abundance.     
 
The dwarf stars were then compared to curves of clusters with known age as seen in Figure\,\ref{fig:Li}\hspace{-2mm}, based on their equivalent width of the Li\,(6708\,\AA) line and their effective temperature. These estimated ages from lithium measurements can be seen as upper limit and classify HIP\,28469 (\#\,1), HIP\,113174 (\#\,2), HIP\,26690 (\#\,3), HIP\,22524 (\#\,5), HIP\,51386 (\#\,7), HIP\,30030 (\#\,9), HIP\,16563 (\#\,13) and HIP\,71631 (\#\,14) as young ($\leq50$\,Myr) according to \cite{tetzlaff}. For HIP\,22524 (\#\,5) we can derive an age $>10$\,Myr, $<5$\,Gyr from HRD isochrone fitting and $\leq50$\,Myr from the lithium test. These ages are consistent with each other and also agree with the estimation of $17.1\pm3.0$\,Myr from \cite{tetzlaff}, as given in Table\,\ref{clearname}\hspace{-2mm}.  
   
For several targets, such as HIP\,2710 (\#\,4), HIP\,44212 (\#\,8), HIP\,19855 (\#\,12), HIP\,7576 (\#\,15) and HIP\,40774 (\#\,16), we could not confirm their ages from \cite{tetzlaff}, because they showed less lithium than we would expect for comparable stars with the same SpT. Within its $2\,\sigma$ uncertainty in Figure\,\ref{fig:Li}\hspace{-2mm}, HIP\,44212 (\#\,8) would cross the 4\,Gyr curve and its age range would therefore be comparable with ages from \cite{ramirez} ($4.98_{-1.90}^{+1.15}$\,Gyr), \cite{casagrande} ($3.78_{-2.34}^{+3.12}$\,Gyr), \cite{pace} ($3.91\pm2.79$\,Gyr) and \cite{mints} ($4.15_{-2.25}^{+4.93}$\,Gyr). Additionally, this star is also located on the 5\,Gyr isochrone in Figure\,\ref{fig:HRD}\hspace{-2mm}. As counterparts to those mentioned stars, HIP\,54531 (\#\,6), HIP\,114385 (\#\,10) and HIP\,115527 (\#\,11) would just barely be, within their $2\,\sigma$ uncertainties, consistent with the young ages from \cite{tetzlaff}.

\begin{figure}[h!]
\centering\includegraphics[width=8.75cm,height=8.75cm,keepaspectratio]{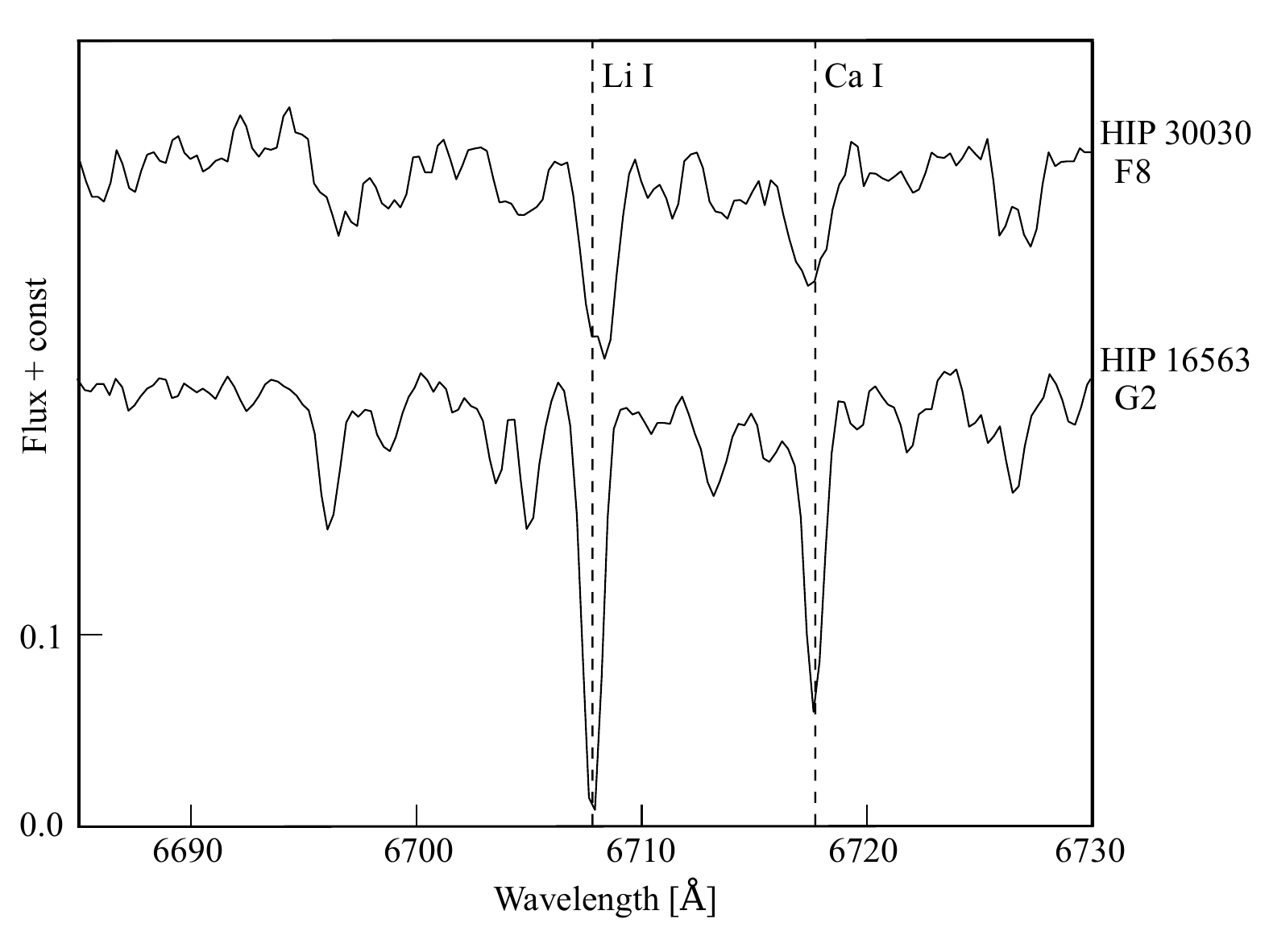}
\caption{Spectra of HIP\,30030 (\#\,9) and HIP\,16563 (\#\,13).}
\label{fig:Li9_13}
\end{figure}

We found two young targets within our sample, namely HIP\,30030 (\#\,9) and HIP\,16563 (\#\,13), that show a relatively large amount of lithium in comparison to their expected lower age limit from their location in the HRD. Their spectra are given in Figure\,\ref{fig:Li9_13}\hspace{-2mm}. These objects were assigned from HRD isochrone fitting to be $\ge20$\,Myr and $>20$\,Myr, respectively. Their position in Figure\,\ref{fig:Li} could suggest an age of $\sim 5$\,Myr. However, these two stars should be handled with care, because as mentioned above the lithium method does not give very reliable age estimations below 20\,Myr \citep{soderblom2014}. Therefore, in combination with isochrone fitting and the lithium test, HIP\,30030 (\#\,9) is more likely older or equal than $20$\,Myr and younger than $50$\,Myr, while HIP\,16563 (\#\,13) is older than $20$\,Myr and younger than $50$\,Myr. \\  

GJ\,182 (HIP\,23200) is one of the youngest stars \citep[e.g.][]{bischoff2020} and, given its distance of $24.38\pm0.02$\,pc \citep{bailerjones}, there is no star known that is both younger and more nearby. Its age was estimated by \cite{bischoff2020} to be ranging between 20\,Myr and 50\,Myr, which is consistent with ages from \cite{brandt}, \cite{binks} and \cite{bell} - and also with membership to the $\beta$\,Pic moving group \citep{lee}. HIP\,113174 (\#\,2) and HIP\,51386 (\#\,7) have comparable ages to GJ\,182 and are not much further away than GJ\,182 ($d\le 2\cdot d_{\text{GJ\,182}}$, as listed in Table\,\ref{clearname}\hspace{-2mm}). Therefore, these two young runaway star candidates, which were listed as field stars in \cite{david2016} and/or \cite{pace}, are the best targets for follow-up investigations of their origin from either dynamical or supernova ejection based on their young age and proximity to the Earth. Even if HIP\,26690 (\#\,3) is further out, given its distance of $167\pm2$\,pc \citep{bailerjones}, it has a comparable age to HIP\,113174 (\#\,2) and HIP\,51386 (\#\,7) and is therefore also a good candidate. In contrast to the mentioned three targets, HIP\,28469 (\#1), HIP\,22524 (\#\,5), HIP\,30030 (\#\,9), HIP\,16563 (\#\,13) and HIP\,71631 (\#\,14) are rather no runaway stars, because they are associated with young nearby stellar clusters \citep{gagne,montes,lopez}. HIP\,28469 (\#1) and HIP\,22524 (\#\,5) were assigned to be members of the Hyades cluster and HIP\,30030 (\#\,9) belongs to the Columba association \citep{gagne}. Furthermore, HIP\,16563 (\#\,13) is part of the AB\,Doradus moving group \citep{gagne} and HIP\,71631 (\#\,14) is listed as member of the Local Association subgroup B4 in \cite{lopez}. As presented in Table\,\ref{clearname}\hspace{-2mm}, our derived age for HIP\,30030 (\#\,9) is consistent with $42_{-4}^{+6}$\,Myr, the age of its associated cluster in \cite{gagne}. We derived an age of 20\,...\,50\,Myr for HIP\,16563 (\#\,13), which younger than $149_{-19}^{+51}$\,Myr of its moving group as given in \cite{gagne}. However, \cite{lopez} list ages for the AB\,Doradus moving group ranging between 30\,Myr and 150\,Myr, which are consistent with our derived age for its possible member star HIP\,16563 (\#\,13). The Local Association subgroup B4 has an average age of $\sim150$\,Myr but also contains stars which are consistent with 80\,Myr \citep{lopez}. That fits with our derived age of 20\,...\,90\,Myr for HIP\,71631 (\#\,14). The determined age of the stars in the Hyades cluster is $750\pm100$\,Myr according to \cite{gagne}. However, the basis for the 625\,Myr isochrone in Figure\,\ref{fig:Li} is also the distribution of the Hyades cluster. Those stars scatter around the 625\,Myr isochrone in the original plot\footnote{\url{http://www.pas.rochester.edu/~emamajek/images/li.jpg}}, that was done by Eric Mamajek, and the location of the lithium richest members are consistent with the location of HIP\,22524 (\#\,5). HIP\,28469 (\#1) is also close to this area within its uncertainties. If HIP\,28469 (\#1) and HIP\,22524 (\#\,5) are actually members of the Hyades cluster, their upper age limit would be then the cluster age.\\ 

\section{Conclusions}\label{sec7}

We carried out spectroscopic follow-up observations for 51 targets from the catalogue by \cite{tetzlaff} to search for the Li\,(6708\,\AA) absorption line, which is a youth indicator. 21 stars have a significantly detected lithium line within their spectra. In combination with isochrones based on the \textit{Gaia}\,DR2, we classified 8 objects as young with ages $\leq50$\,Myr. Some of these targets are already associated with young nearby stellar clusters. HIP\,113174 (\#\,2), HIP\,26690 (\#\,3) and HIP\,51386 (\#\,7) are the remaining young runaway star candidates, which are outside of known clusters. They are suitable for further follow-up observations to identify their place of origin and/or to search for possible companions.
	
As it is the standard in our survey the fully reduced FLECHAS spectra as well as the measured equivalent widths of the Li\,(6708\,\AA) line will be made available in \texttt{VizieR} after publication. 

\begin{table}[h!]
\caption{Radial velocity measurements for the spectroscopic binaries as derived from the Ca\,(6718\,\AA) line and given with their Barycentric Julian Date (BJD).}
\resizebox{0.5\textwidth}{!}{
\centering
\begin{tabular}{lcrr}
\hline
Target  & BJD  & $RV_{1}$ [km/s]  & $RV_{2}$ [km/s]     \\
\hline
HIP\,5081   & 2459061.55093 & $54.4\pm 4.0$ & $-50.9\pm4.1$ \\
HIP\,22524  & 2458913.32631 & $32.2\pm 2.9$ &               \\
HIP\,26690  & 2458933.32562 & $19.4\pm 3.4$ &               \\
HIP\,64312  & 2458912.45917 & $-4.5\pm 3.3$ &               \\
HIP\,71631  & 2458927.37139 & $-22.2\pm1.6$ &               \\
HIP\,82798  & 2458912.58774 & $-30.1\pm3.0$ &               \\
HIP\,112821 & 2459044.49852 & $-4.4\pm 2.2$ &               \\
HIP\,114379 & 2459052.52385 & $15.7\pm 1.6$ & $-32.5\pm2.5$ \\
\hline
\end{tabular}
}                                              		
\label{RV}
\end{table}         

\section*{Acknowledgments}

We thank additional observers who have been involved in some of the observations of this project, obtained at the University Observatory Jena and the Fred L.\ Whipple Observatory, in particular P. Berlind, S. Bischoff, M. Calkins, G. Esquerdo, L. Nueva and S. Quinn.
This publication makes use of data products of the \texttt{VizieR} databases, operated at CDS, Strasbourg, France. We also thank the \textit{Gaia} Data Processing and Analysis Consortium of the European Space Agency (ESA) for processing and providing the data of the \textit{Gaia} mission.\\
We thank Eric Mamajek for providing the curves of lithium as age indicator for F\,G\,K\,M stars.\\
This work was supported by the Deutsche Forschungsgemeinschaft with financing the projects \fundingNumber{NE 515/58-1} and \fundingNumber{MU 2695/27-1}.\\
We thank the referee for helpful comments, which improved our manuscript.

\appendix\newpage

\section{Additional target information\label{app1}}

\begin{center}
\vspace{0.8cm}
\tablefirsthead{%
\hline
Target          & \multicolumn{1}{c}{Date [UT]} 	& \multicolumn{1}{c}{T$_{\text{exp}}$ [s]}  & \multicolumn{1}{c}{SNR} \\
\hline}
\tablehead{%
\multicolumn{1}{l}{Continued}\\
\hline
Target          & \multicolumn{1}{c}{Date [UT]} 	& \multicolumn{1}{c}{T$_{\text{exp}}$ [s]}  & \multicolumn{1}{c}{SNR} \\
				\hline}
\tabletail{%
\hline	}
\tablelasttail{\hline}
\tablecaption{Observation log. For each target we list the date and mid-time of the observation,  as well as the individual exposure time (T$_{\text{exp}}$) of each FLECHAS spectrum and the signal-to-noise ratios (SNR) of the average of the three spectra, as measured at $\lambda = 6700$\,\AA.}
\begin{supertabular}{cccc}
HIP\,1762   & 2020 Mar 23, 19:42 &  900  & 76   \\
HIP\,2710   & 2020 Sep 03, 01:02 &  450  & 101   \\
HIP\,5081   & 2020 Jul 31, 01:10 &  150  & 102   \\
HIP\,5913   & 2020 Mar 09, 21:10 &  900  & 93   \\
HIP\,7576   & 2020 Sep 03, 01:36 &  900  & 111   \\
HIP\,10141  & 2020 Mar 03, 23:16 &  600  & 78   \\
HIP\,11126  & 2020 Aug 05, 01:37 &  1200 & 87   \\
HIP\,11429  & 2020 Mar 04, 23:22 &  300  & 108   \\
HIP\,12083  & 2020 Mar 03, 22:16 &  600  & 98   \\
HIP\,12297  & 2020 Mar 21, 22:10 &  300  & 82   \\
HIP\,16563  & 2020 Mar 23, 20:36 &  900  & 91   \\
HIP\,19587  & 2020 Sep 03, 03:18 &  150  & 201   \\
HIP\,19855  & 2020 Mar 18, 19:23 &  450  & 103   \\
HIP\,22524  & 2020 Mar 04, 19:50 &  450  & 118   \\
HIP\,24817  & 2020 Mar 04, 19:32 &  150  & 160   \\
HIP\,25749  & 2020 Mar 04, 00:16 &  450  & 94   \\
HIP\,26690  & 2020 Mar 24, 19:50 &  1200 & 74   \\
HIP\,27841  & 2020 Mar 22, 20:16 &  600  & 76   \\
HIP\,28469  & 2020 Mar 18, 19:52 &  600  & 101   \\
HIP\,30030  & 2020 Mar 22, 19:32 &  900  & 50   \\
HIP\,35219  & 2020 Mar 03, 22:35 &  300  & 99   \\
HIP\,40774  & 2020 Mar 23, 21:42 &  1200 & 93   \\
HIP\,44212  & 2020 Mar 22, 21:31 &  900  & 87   \\
HIP\,47725  & 2020 Mar 21, 21:19 &  900  & 90   \\
HIP\,49700  & 2020 Mar 09, 23:18 &  900  & 121   \\
HIP\,51386  & 2020 Mar 04, 01:01 &  450  & 117   \\
HIP\,54531  & 2020 Mar 22, 22:42 &  1200 & 81   \\
HIP\,56770  & 2020 Mar 03, 21:14 &  150  & 123   \\
HIP\,57160  & 2020 Mar 10, 00:09 &  900  & 126   \\
HIP\,57529  & 2020 Mar 10, 01:19 &  600  & 126   \\
HIP\,64312  & 2020 Mar 03, 22:54 &  300  & 89   \\
HIP\,71631  & 2020 Mar 18, 20:52 &  600  & 90   \\
HIP\,74333  & 2020 Mar 04, 01:38 &  450  & 100   \\
HIP\,82798  & 2020 Mar 04, 02:06 &  300  & 102   \\
HIP\,89828  & 2020 Jul 30, 21:42 &  600  & 92   \\
HIP\,91594  & 2020 Mar 04, 04:35 &  450  & 84   \\
HIP\,103471 & 2020 Mar 04, 03:11 &  450  & 155   \\
HIP\,105607 & 2020 Jul 06, 22:47 &  300  & 83   \\
HIP\,106053 & 2020 Jul 06, 23:20 &  300  & 71   \\
HIP\,111446 & 2020 Jul 30, 02:05 &  600  & 91   \\
HIP\,111647 & 2020 Aug 07, 01:21 &  600  & 85   \\
HIP\,112821 & 2020 Jul 13, 23:54 &  600  & 83   \\
HIP\,113174 & 2020 Jul 06, 22:29 &  150  & 101   \\
HIP\,113811 & 2020 Jul 31, 00:33 &  900  & 121   \\
HIP\,113952 & 2020 Jul 22, 00:00 &  300  & 85   \\
HIP\,114379 & 2020 Jul 22, 00:32 &  900  & 124   \\
HIP\,114385 & 2000 Jul 21, 23:34 &  450  & 103   \\
HIP\,115527 & 2020 Aug 07, 02:21 &  600  & 91   \\
HIP\,115906 & 2020 Jul 31, 01:45 &  300  & 168   \\
HIP\,116983 & 2020 Jul 29, 23:42 &  1200 & 83   \\
HIP\,117835 & 2020 Jul 30, 00:59 &  900  & 89   \\
\end{supertabular}
\label{tab:obslog}
\end{center}

\newpage

\begin{center}
\vspace{0.8cm}
\tablefirsthead{%
\hline
Target          & \multicolumn{1}{c}{$EW_{\text{Li}}$ [m\AA]} &  \multicolumn{1}{c}{Target} &	 \multicolumn{1}{c}{ $EW_{\text{Li}}$ [m\AA]} \\
\hline}
\tablehead{%
\multicolumn{1}{l}{Continued}\\
\hline
Target          & \multicolumn{1}{c}{$EW_{\text{Li}}$ [m\AA]} &  \multicolumn{1}{c}{Target} &	 \multicolumn{1}{c}{ $EW_{\text{Li}}$ [m\AA]} \\ 	
				\hline}
\tabletail{%
\hline	}
\tablelasttail{\hline}
\tablecaption{Measured equivalent widths ($EW_{\text{Li}}$) of the Li\,(6708\,\AA) line. Targets without significant detection ($EW_{\text{Li}}< 3\cdot\sigma EW_{\text{Li}}$) are listed with "-".}
\begin{supertabular}{cccc}
HIP\,1762   & $\,\,~~55\pm14$   & HIP\,54531  & $\,\,~~85\pm14$   \\
HIP\,2710   & $\,\,~~36\pm12$   & HIP\,56770  & $~~~-  $    \\
HIP\,5081   & $~~~-  $          & HIP\,57160  & $~~~-  $    \\
HIP\,5913   & $~~~-  $          & HIP\,57529  & $~~~-  $    \\
HIP\,7576   & $~114\pm12$        & HIP\,64312  & $~~~-  $    \\
HIP\,10141  & $~~~-  $          & HIP\,71631  & $~197\pm13$  \\
HIP\,11126  & $~~~-  $          & HIP\,74333  & $~~~-  $    \\
HIP\,11429  & $~~~-  $          & HIP\,82798  & $\,\,~~37\pm11$   \\
HIP\,12083  & $\,\,~~98\pm12$   & HIP\,89828  & $~~~-  $    \\
HIP\,12297  & $~~~-  $          & HIP\,91594  & $~~~-  $    \\
HIP\,16563  & $~254\pm14$       & HIP\,103471 & $~~95\pm8$    \\
HIP\,19587  & $~~~-  $          & HIP\,105607 & $~~~-  $    \\
HIP\,19855  & $\,\,~~71\pm12$   & HIP\,106053 & $\,\,~~76\pm16$   \\
HIP\,22524  & $~119\pm12$       & HIP\,111446 & $~~~-  $    \\
HIP\,24817  & $~~~-  $          & HIP\,111647 & $~~~-  $    \\
HIP\,25749  & $~~~-  $          & HIP\,112821 & $~~~-  $    \\
HIP\,26690  & $\,\,~~66\pm15$   & HIP\,113174 & $\,\,~~57\pm11$   \\
HIP\,27841  & $~~~-  $          & HIP\,113811 & $~~~-  $    \\
HIP\,28469  & $\,\,~~61\pm12$   & HIP\,113952 & $~~~-  $    \\
HIP\,30030  & $~190\pm24$       & HIP\,114379 & $~~~-  $    \\
HIP\,35219  & $~~~-  $          & HIP\,114385 & $~111\pm10$  \\
HIP\,40774  & $~119\pm12$       & HIP\,115527 & $~126\pm13$  \\
HIP\,44212  & $\,\,~~64\pm13$   & HIP\,115906 & $~~~-  $    \\
HIP\,47725  & $~~~-  $          & HIP\,116983 & $~~~-  $    \\
HIP\,49700  & $~~~-  $          & HIP\,117835 & $~~~-  $    \\
HIP\,51386  & $~120\pm11$       &            &               \\
\end{supertabular}
\label{tab:Li_all}
\end{center}

\newpage
\onecolumn

\begin{center}
\vspace{0.8cm}
\tablefirsthead{%
	\hline
	Target          & \multicolumn{1}{c}{$d$ [pc]} &  \multicolumn{1}{c}{$G$ [mag]} &	 \multicolumn{1}{c}{$A_{\text{G}}$ [mag]}  & \multicolumn{1}{c}{\textit{$M_{\text{G}}$} [mag]}  & \multicolumn{1}{c}{$T_{\text{eff}}$ [K] }  & \multicolumn{1}{c}{$R$ [R$_{\odot}$]}  & \multicolumn{1}{c}{$L$ [L$_{\odot}$] } &\multicolumn{1}{c}{rem.}\\
	\hline}
\tablehead{%
\multicolumn{1}{l}{Continued}\\
\hline
	Target          & \multicolumn{1}{c}{$d$ [pc]} &  \multicolumn{1}{c}{$G$ [mag]} &	 \multicolumn{1}{c}{$A_{\text{G}}$ [mag]}  & \multicolumn{1}{c}{\textit{$M_{\text{G}}$} [mag]}  & \multicolumn{1}{c}{$T_{\text{eff}}$ [K] }  & \multicolumn{1}{c}{$R$ [R$_{\odot}$]}  & \multicolumn{1}{c}{$L$ [L$_{\odot}$] } &\multicolumn{1}{c}{rem.}\\
	\hline}
\tabletail{%
	\hline	}
\tablelasttail{
	\hline}
\tablecaption{Physical properties of the targets, namely the corrected apparent brightness $G$ according to \cite{weiler}, the effective temperature $T_{\text{eff}}$, stellar radius $R$, and luminosity $L$ of the target stars (if available), from the \textit{Gaia}\,DR2, as well as the derived extinction in the $G$-band $A_{\text{G}}$ from \cite{gontcharov}. From these parameters together with the distances $d$ by \cite{bailerjones}, the absolute $G$-band brightness $M_{\text{G}}$ of the targets was calculated. The last column yields further remarks, with details listed in the footnote of this table.}
\begin{supertabular}{ccccccccc}
HIP\,1762   & $217_{-2  }^{+2}   $ & $7.928\pm0.001$ & $0.245_{-0.099}^{+0.100}$ & $1.000_{-0.120}^{+0.118}$  & $6454_{-116}^{+136}$ & $3.86_{-0.15}^{+0.15}$ & $23.3_{-0.3}^{+0.3     }$ &   \\
HIP\,2710   & $40.6_{-0.1}^{+0.1}$ & $6.782\pm0.001$ & $0.087_{-0.087}^{+0.099}$ & $3.654_{-0.104}^{+0.091}$  & $6372_{-81 }^{+129}$ & $1.25_{-0.05}^{+0.03}$ & $2.33_{-0.01}^{+0.01   }$ &   \\
HIP\,5081   & $52.5_{-0.4}^{+0.4}$ & $5.525\pm0.002$ & $0.079_{-0.079}^{+0.099}$ & $2.375_{-0.117}^{+0.097}$  & $6567_{-104}^{+72 }$ & $-					$ & $-$ 					&  b \\
HIP\,5913   & $58.6_{-0.2}^{+0.2}$ & $7.202\pm0.001$ & $0.126_{-0.098}^{+0.099}$ & $3.236_{-0.105}^{+0.104}$  & $6605_{-226}^{+134}$ & $1.39_{-0.06}^{+0.10}$ & $3.30_{-0.01}^{+0.01   }$ &   \\
HIP\,7576   & $24.0_{-0.1}^{+0.1}$ & $7.442\pm0.001$ & $0.047_{-0.047}^{+0.099}$ & $5.491_{-0.101}^{+0.050}$  & $5303_{-71 }^{+73 }$ & $0.82_{-0.03}^{+0.02}$ & $0.474_{-0.001}^{+0.001}$ &   \\
HIP\,10141  & $99.2_{-0.4}^{+0.4}$ & $6.559\pm0.001$ & $0.150_{-0.098}^{+0.099}$ & $1.428_{-0.109}^{+0.107}$  & $7077_{-233}^{+165}$ & $2.76_{-0.12}^{+0.19}$ & $17.2_{-0.1}^{+0.1     }$ &   \\
HIP\,11126  & $180_{-2}^{+2}     $ & $8.056\pm0.001$ & $0.205_{-0.099}^{+0.100}$ & $1.574_{-0.120}^{+0.119}$  & $6380_{-134}^{+122}$ & $3.09_{-0.12}^{+0.13}$ & $14.2_{-0.2}^{+0.2     }$ &   \\
HIP\,11429  & $93.3_{-0.4}^{+0.4}$ & $7.341\pm0.001$ & $0.134_{-0.098}^{+0.099}$ & $2.357_{-0.109}^{+0.108}$  & $6716_{-81 }^{+46 }$ & $2.00_{-0.02}^{+0.05}$ & $7.36_{-0.04}^{+0.04   }$ &   \\
HIP\,12083  & $197_{-1}^{+1}     $ & $6.529\pm0.001$ & $0.260_{-0.099}^{+0.100}$ &$-0.202_{-0.114}^{+0.113}$  & $6562_{-103}^{+115}$ & $6.46_{-0.22}^{+0.21}$ & $69.8_{-0.7}^{+0.7     }$ &   \\
HIP\,12297  & $179_{-2}^{+3}     $ & $7.552\pm0.001$ & $0.323_{-0.099}^{+0.101}$ & $0.968_{-0.131}^{+0.129}$  & $6702_{-167}^{+157}$ & $3.50_{-0.16}^{+0.18}$ & $22.3_{-0.4}^{+0.4     }$ &   \\
HIP\,16563  & $36.4_{-0.1}^{+0.1}$ & $7.969\pm0.002$ & $0.118_{-0.098}^{+0.099}$ & $5.046_{-0.106}^{+0.104}$  & $5612_{-153}^{+176}$ & $0.85_{-0.05}^{+0.05}$ & $0.645_{-0.002}^{+0.002}$ &   \\
HIP\,19587  & $36.5_{-0.6}^{+0.6}$ & $3.946\pm0.003$ & $0.110_{-0.098}^{+0.099}$ & $1.023_{-0.136}^{+0.134}$  & $6930_{-320}^{+120}$ & $3.62_{-0.13}^{+0.35}$ & $27.2_{-0.5}^{+0.5     }$ &   \\
HIP\,19855  & $22.1_{-0.1}^{+0.1}$ & $6.760\pm0.001$ & $0.063_{-0.063}^{+0.099}$ & $4.976_{-0.102}^{+0.066}$  & $5648_{-39 }^{+108}$ & $0.89_{-0.03}^{+0.01}$ & $0.724_{-0.001}^{+0.001}$ &   \\
HIP\,22524  & $50.4_{-0.1}^{+0.1}$ & $7.148\pm0.001$ & $0.158_{-0.098}^{+0.099}$ & $3.477_{-0.379}^{+0.105}$  & $6246_{-60 }^{+86 }$ & $-					$ & $-						$ & b  \\
HIP\,24817  & $61.3_{-0.6}^{+0.6}$ & $5.232\pm0.003$ & $0.181_{-0.098}^{+0.100}$ & $1.112_{-0.123}^{+0.121}$  & $6602_{-114}^{+97 }$ & $3.64_{-0.11}^{+0.13}$ & $22.7_{-0.3}^{+0.3     }$ &   \\
HIP\,25749  & $55.1_{-0.1}^{+0.1}$ & $7.026\pm0.001$ & $0.134_{-0.098}^{+0.099}$ & $3.185_{-0.105}^{+0.103}$  & $6663_{-136}^{+80 }$ & $1.39_{-0.03}^{+0.06}$ & $3.43_{-0.01}^{+0.01   }$ &   \\
HIP\,26690  & $167_{-2}^{+2}     $ & $8.302\pm0.001$ & $0.237_{-0.099}^{+0.100}$ & $2.555_{-0.166}^{+0.171}$  & $6659_{-128}^{+67 }$ & $-					$ & $-						$ & b  \\
HIP\,27841  & $301_{-5}^{+5}     $ & $7.453\pm0.001$ & $0.284_{-0.099}^{+0.100}$ &$-0.222_{-0.134}^{+0.132}$  & $6166_{-121}^{+139}$ & $7.33_{-0.32}^{+0.30}$ & $70.0_{-1.5}^{+1.5     }$ &   \\
HIP\,28469  & $91.1_{-0.4}^{+0.4}$ & $7.400\pm0.001$ & $0.166_{-0.098}^{+0.100}$ & $2.436_{-0.110}^{+0.108}$  & $6701_{-87 }^{+150}$ & $1.91_{-0.08}^{+0.05}$ & $6.65_{-0.04}^{+0.04   }$ &   \\
HIP\,30030  & $51.9_{-0.1}^{+0.1}$ & $7.697\pm0.002$ & $0.213_{-0.099}^{+0.100}$ & $3.906_{-0.108}^{+0.107}$  & $6027_{-69 }^{+36 }$ & $1.18_{-0.01}^{+0.03}$ & $1.65_{-0.01}^{+0.01   }$ &   \\
HIP\,35219  & $52.2_{-1.0}^{+1.0}$ & $6.628\pm0.001$ & $0.174_{-0.098}^{+0.100}$ & $2.866_{-0.141}^{+0.139}$  & $6511_{-54 }^{+79 }$ & $1.65_{-0.03}^{+0.03}$ & $4.44_{-0.09}^{+0.09   }$ &   \\
HIP\,40774  & $22.4_{-0.1}^{+0.1}$ & $8.069\pm0.001$ & $0.071_{-0.071}^{+0.099}$ & $6.249_{-0.109}^{+0.081}$  & $4917_{-58 }^{+67 }$ & $0.69_{-0.02}^{+0.01}$ & $0.247_{-0.001}^{+0.001}$ &   \\
HIP\,44212  & $46.2_{-0.1}^{+0.1}$ & $7.646\pm0.001$ & $0.166_{-0.098}^{+0.100}$ & $4.158_{-0.105}^{+0.104}$  & $6041_{-74 }^{+44 }$ & $1.07_{-0.02}^{+0.02}$ & $1.37_{-0.00}^{+0.00   }$ &   \\
HIP\,47725  & $72.1_{-0.2}^{+0.2}$ & $7.590\pm0.001$ & $0.047_{-0.047}^{+0.099}$ & $3.252_{-0.105}^{+0.054}$  & $6762_{-237}^{+307}$ & $1.36_{-0.11}^{+0.10}$ & $3.49_{-0.01}^{+0.01   }$ &   \\
HIP\,49700  & $80.0_{-0.4}^{+0.4}$ & $7.708\pm0.001$ & $0.095_{-0.095}^{+0.099}$ & $3.097_{-0.109}^{+0.105}$  & $6575_{-134}^{+152}$ & $1.51_{-0.06}^{+0.07}$ & $3.85_{-0.02}^{+0.02   }$ &   \\
HIP\,51386  & $31.0_{-0.1}^{+0.1}$ & $6.711\pm0.001$ & $0.055_{-0.055}^{+0.099}$ & $4.198_{-0.106}^{+0.062}$  & $6111_{-21 }^{+147}$ & $1.08_{-0.05}^{+0.00}$ & $1.46_{-0.00}^{+0.01   }$ &   \\
HIP\,54531  & $61.0_{-0.2}^{+0.2}$ & $8.104\pm0.001$ & $0.095_{-0.095}^{+0.099}$ & $4.082_{-0.107}^{+0.103}$  & $6218_{-114}^{+63 }$ & $1.08_{-0.03}^{+0.04}$ & $1.56_{-0.01}^{+0.01   }$ &   \\
HIP\,56770  & $47.8_{-0.5}^{+0.5}$ & $5.459\pm0.002$ & $0.008_{-0.008}^{+0.099}$ & $2.054_{-0.124}^{+0.033}$  & $6934_{-156}^{+74 }$ & $2.31_{-0.05}^{+0.11}$ & $11.1_{-0.1}^{+0.1     }$ &   \\
HIP\,57160  & $147_{-1}^{+1}     $ & $7.579\pm0.001$ & $0.087_{-0.087}^{+0.099}$ & $1.657_{-0.113}^{+0.100}$  & $6830_{-260}^{+127}$ & $2.74_{-0.10}^{+0.22}$ & $14.7_{-0.1}^{+0.1     }$ &   \\
HIP\,57529  & $92.5_{-5.6}^{+6.3}$ & $7.204\pm0.005$ & $0.047_{-0.047}^{+0.099}$ & $2.326_{-0.246}^{+0.186}$  & $6531_{-218}^{+154}$ & $-                   $ & $-                      $ & a   \\
HIP\,64312  & $156_{-3}^{+4}     $ & $6.724\pm0.001$ & $0.189_{-0.098}^{+0.100}$ & $0.805_{-0.150}^{+0.148}$  & $6742_{-87 }^{+174}$ & $-					$ & $-						$ & b  \\
HIP\,71631  & $34.4_{-0.1}^{+0.1}$ & $7.492\pm0.001$ & $0.039_{-0.039}^{+0.099}$ & $4.843_{-0.118}^{+0.115}$  & $5584_{-193}^{+115}$ & $-					$ & $-						$ & b  \\
HIP\,74333  & $155_{-1}^{+1}     $ & $7.060\pm0.001$ & $0.118_{-0.098}^{+0.099}$ & $0.990_{-0.116}^{+0.114}$  & $7089_{-81 }^{+211}$ & $3.42_{-0.20}^{+0.08}$ & $26.6_{-0.3}^{+0.3     }$ &   \\
HIP\,82798  & $75.4_{-0.3}^{+0.3}$ & $6.260\pm0.001$ & $0.181_{-0.098}^{+0.100}$ & $1.691_{-0.380}^{+0.107}$  & $7100_{-124}^{+160}$ & $-					$ & $-						$ & b  \\
HIP\,89828  & $98.8_{-0.6}^{+0.6}$ & $7.231\pm0.001$ & $0.395_{-0.100}^{+0.101}$ & $1.861_{-0.115}^{+0.113}$  & $6060_{-177}^{+364}$ & $2.75_{-0.30}^{+0.17}$ & $9.22_{-0.07}^{+0.07   }$ &   \\
HIP\,91594  & $133_{-2}^{+2}     $ & $7.124\pm0.001$ & $0.402_{-0.100}^{+0.101}$ & $1.103_{-0.131}^{+0.130}$  & $6911_{-346}^{+335}$ & $2.99_{-0.27}^{+0.32}$ & $18.3_{-0.3}^{+0.3     }$ &   \\
HIP\,103471 & $185_{-1}^{+1}     $ & $7.180\pm0.001$ & $0.142_{-0.098}^{+0.099}$ & $0.707_{-0.113}^{+0.112}$  & $7007_{-263}^{+144}$ & $3.94_{-0.15}^{+0.31}$ & $33.7_{-0.3}^{+0.3     }$ &   \\
HIP\,105607 & $109_{-1}^{+1}     $ & $6.421\pm0.001$ & $0.197_{-0.098}^{+0.100}$ & $1.044_{-0.120}^{+0.119}$  & $6241_{-47 }^{+16 }$ & $4.14_{-0.02}^{+0.06}$ & $23.4_{-0.1}^{+0.1     }$ &   \\
HIP\,106053 & $105_{-9}^{+11}    $ & $6.546\pm0.008$ & $0.118_{-0.098}^{+0.099}$ & $1.320_{-0.326}^{+0.305}$  & $6531_{-218}^{+154}$ & $-                   $ & $-                      $ & a  \\
HIP\,111446 & $66.4_{-0.4}^{+0.4}$ & $7.487\pm0.001$ & $0.110_{-0.098}^{+0.099}$ & $3.265_{-0.113}^{+0.112}$  & $6640_{-148}^{+175}$ & $1.36_{-0.07}^{+0.07}$ & $3.25_{-0.02}^{+0.02   }$ &   \\
HIP\,111647 & $84.6_{-0.4}^{+0.4}$ & $7.381\pm0.003$ & $0.118_{-0.098}^{+0.099}$ & $2.626_{-0.111}^{+0.110}$  & $7100_{-144}^{+213}$ & $1.60_{-0.09}^{+0.07}$ & $5.86_{-0.04}^{+0.04   }$ &   \\
HIP\,112821 & $101_{-1}^{+1}     $ & $7.294\pm0.001$ & $0.095_{-0.095}^{+0.099}$ & $2.180_{-0.380}^{+0.106}$  & $5728_{-128}^{+92 }$ & $-					$ & $-						$ & b  \\
HIP\,113174 & $42.8_{-0.1}^{+0.1}$ & $5.799\pm0.001$ & $0.095_{-0.095}^{+0.099}$ & $2.549_{-0.105}^{+0.101}$  & $6684_{-164}^{+159}$ & $1.89_{-0.09}^{+0.10}$ & $6.43_{-0.02}^{+0.02   }$ &   \\
HIP\,113811 & $676_{-290}^{+2027}$ & $7.405\pm0.011$ & $0.147_{-0.049}^{+0.049}$ &$-1.891_{-3.070}^{+1.275}$  & $4457_{-161}^{+297}$ & $-                   $ & $-                      $ & a\,c  \\
HIP\,113952 & $72.1_{-0.2}^{+0.2}$ & $6.651\pm0.001$ & $0.103_{-0.098}^{+0.099}$ & $2.260_{-0.106}^{+0.105}$  & $6680_{-112}^{+301}$ & $2.15_{-0.18}^{+0.07}$ & $8.30_{-0.04}^{+0.04   }$ &   \\
HIP\,114379 & $30.4_{-0.1}^{+0.1}$ & $7.653\pm0.001$ & $0.079_{-0.079}^{+0.099}$ & $5.845_{-0.107}^{+0.087}$  & $4979_{-106}^{+169}$ & $-					$ & $-						$ & b  \\
HIP\,114385 & $30.3_{-0.1}^{+0.1}$ & $6.980\pm0.001$ & $0.079_{-0.079}^{+0.099}$ & $4.493_{-0.106}^{+0.086}$  & $5801_{-41 }^{+103}$ & $1.04_{-0.04}^{+0.01}$ & $1.10_{-0.01}^{+0.01   }$ &   \\
HIP\,115527 & $30.4_{-0.1}^{+0.1}$ & $7.432\pm0.001$ & $0.071_{-0.071}^{+0.099}$ & $4.945_{-0.105}^{+0.077}$  & $5742_{-78 }^{+135}$ & $0.87_{-0.04}^{+0.02}$ & $0.734_{-0.003}^{+0.002}$ &   \\
HIP\,115906 & $287_{-43}^{+62}   $ & $5.680\pm0.005$ & $0.142_{-0.098}^{+0.099}$ &$-1.754_{-0.531}^{+0.459}$  & $4853_{-297}^{+235}$ & $-                   $ & $-                      $ & a  \\
HIP\,116983 & $111_{-1}^{+1}     $ & $8.237\pm0.001$ & $0.071_{-0.071}^{+0.099}$ & $2.932_{-0.111}^{+0.083}$  & $7163_{-253}^{+233}$ & $1.40_{-0.09}^{+0.10}$ & $4.62_{-0.03}^{+0.04   }$ &   \\
HIP\,117835 & $83.7_{-0.3}^{+0.3}$ & $7.903\pm0.001$ & $0.055_{-0.055}^{+0.099}$ & $3.235_{-0.107}^{+0.064}$  & $6694_{-169}^{+213}$ & $1.39_{-0.08}^{+0.08}$ & $3.52_{-0.02}^{+0.02   }$ &   \\
\end{supertabular}
\label{tab:gaia_infos}
\end{center}
\hspace{10pt}a - \textit{Hipparcos} parallax \citep{vanleeuwen} was used for distance determination, $H_{p}$ and $V-I$ magnitudes from \\
\hspace{12pt} \textit{Hipparcos} \citep{perryman} were transformed into $G$-band magnitude according to \cite{evans}, \\
\hspace{12pt} $T_{\text{eff}}$ \citep{damiani} according to the \textit{Hipparcos} spectral type\\
b - spectroscopic binary ($M_{\text{G}}$ therefore up to $\sim0.75$\,mag fainter as expected from \textit{Gaia}\,DR2 entry)\\
c - $E(g-r)$ from \cite{green} was converted into $A_{\text{G}}$ according to \cite{wang}  
\newpage
\twocolumn

\section{Dwarf stars with significant lithium detection}\label{app2}

\includegraphics[width=11cm,height=20.5cm,keepaspectratio]{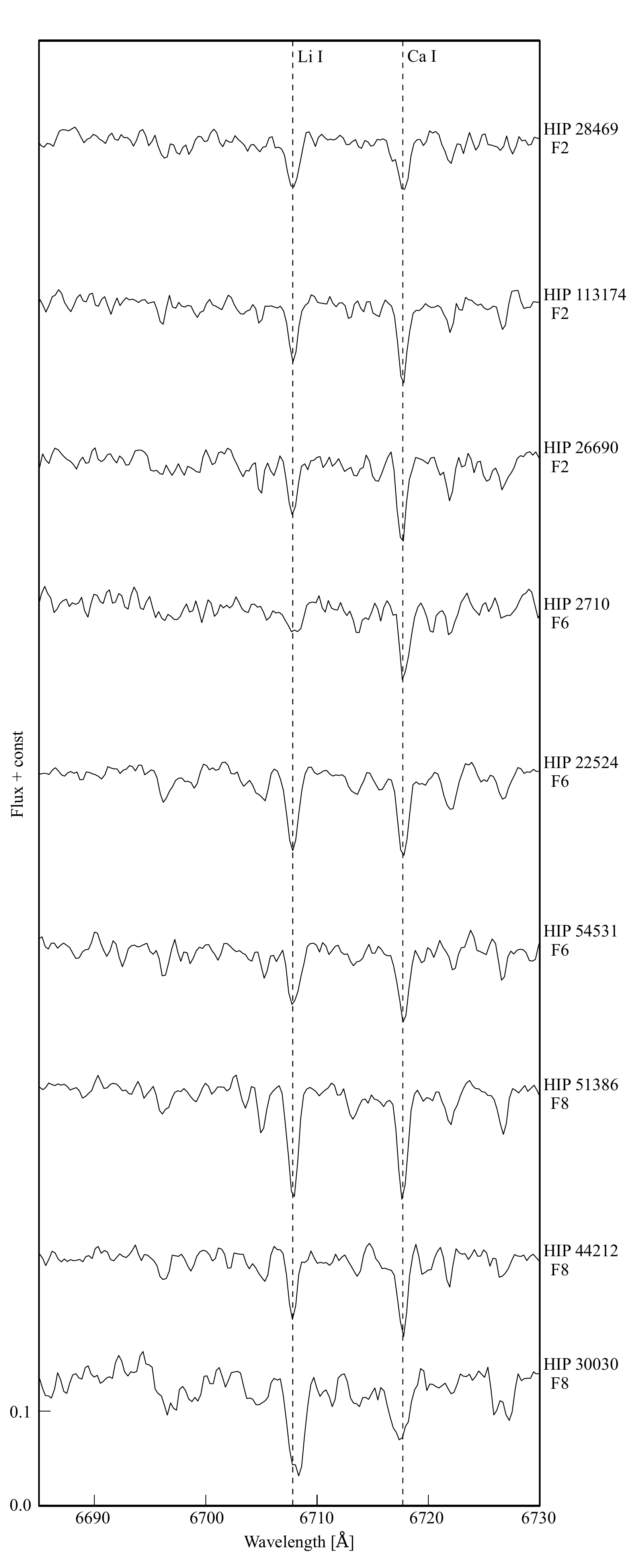}

\newpage
\vspace*{15pt}
\includegraphics[width=11cm,height=20.5cm,keepaspectratio]{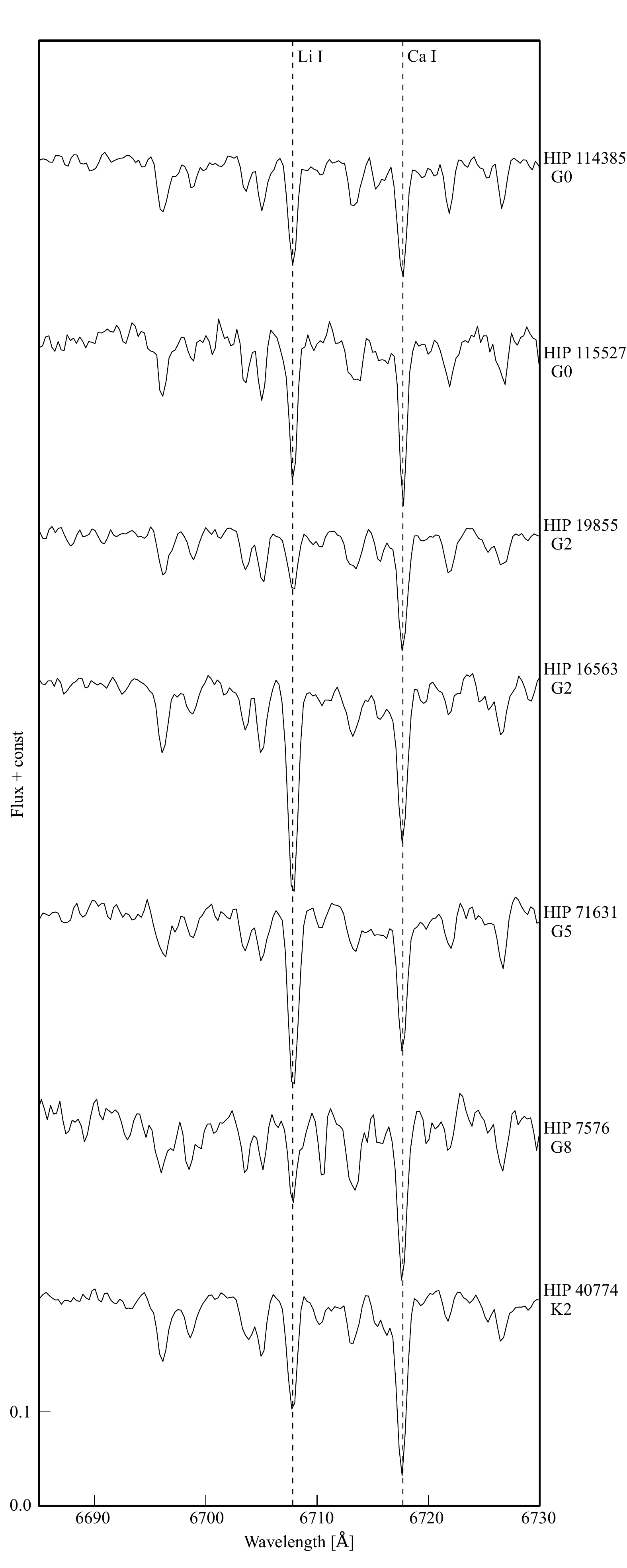}

\newpage

\section{Sub-giant/giant stars with significant lithium detection}\label{app3}

\includegraphics[width=11cm,height=20.5cm,keepaspectratio]{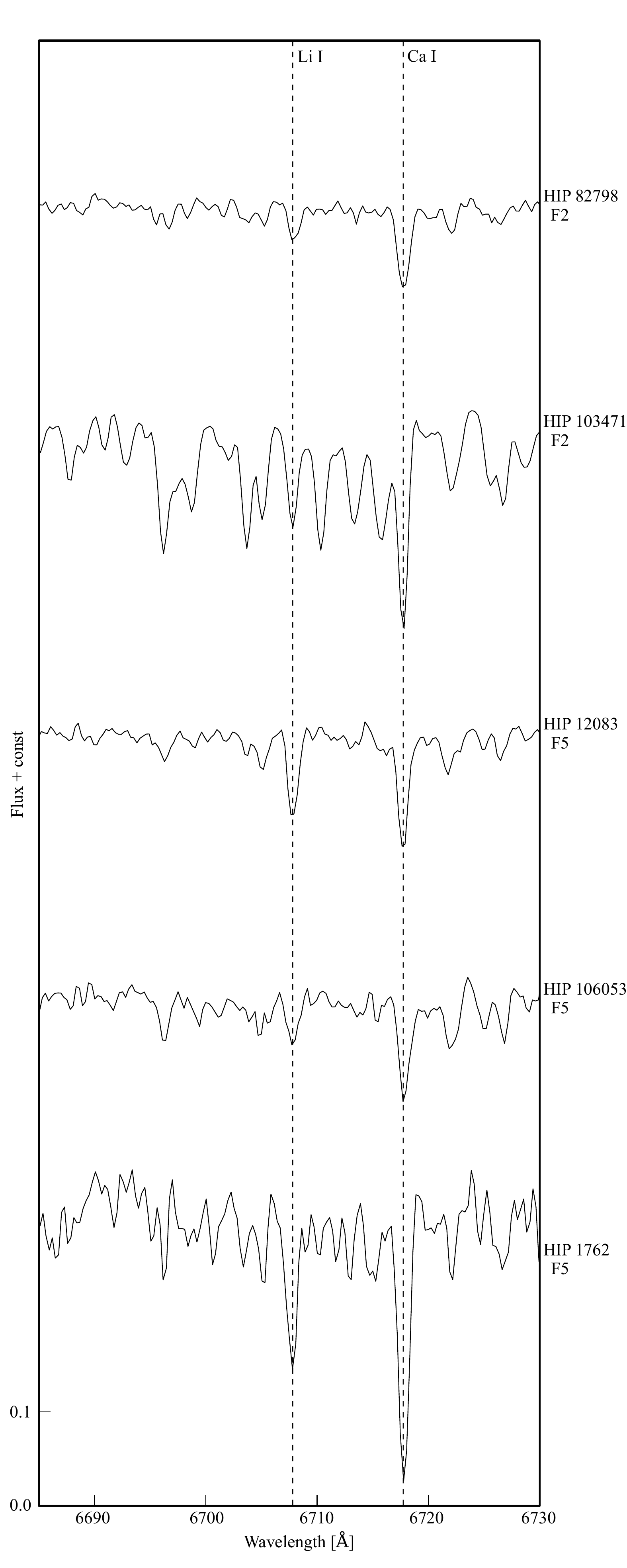}
\newpage

\nocite{*}

\bibliography{bischoff}

\section*{Author Biography}

Richard Bischoff is a PhD student at the Astrophysical Institute and University Observatory Jena. His main field of research are photometry and spectroscopy of exoplanet candidate host stars.

\end{document}